\documentclass[%
 reprint,
 superscriptaddress,
 amsmath,amssymb,
 aps,
]{revtex4-1}

\usepackage[colorlinks, citecolor=red]{hyperref}
\usepackage{graphicx}
\usepackage{dcolumn}
\usepackage{bm}

\usepackage{listings}
\usepackage{color}
\usepackage{subfigure}
\usepackage{algorithm}
\usepackage{algorithmic}
\usepackage{bbm}
\usepackage{ntheorem}

\newtheorem{Theo}{Theorem}

\definecolor{dkgreen}{rgb}{0,0.6,0}
\definecolor{gray}{rgb}{0.5,0.5,0.5}
\definecolor{mauve}{rgb}{0.58,0,0.82}

\begin{document}


\title{Quantum Newton's method for solving system of nonlinear algebraic equations}
\author{Cheng Xue}
 \affiliation{CAS Key Laboratory of Quantum Information, University of Science and Technology of China, Hefei, Anhui 230026, P. R. China}
\author{Yu-Chun Wu}
 \email{wuyuchun@ustc.edu.cn} 
 \affiliation{CAS Key Laboratory of Quantum Information, University of Science and Technology of China, Hefei, Anhui 230026, P. R. China}
 \affiliation{Institute of Artificial Intelligence, Hefei Comprehensive National Science Center Hefei, Anhui 230026, P. R. China}
\author{Guo-Ping Guo}
 \affiliation{CAS Key Laboratory of Quantum Information, University of Science and Technology of China, Hefei, Anhui 230026, P. R. China}
 \affiliation{Institute of Artificial Intelligence, Hefei Comprehensive National Science Center Hefei, Anhui 230026, P. R. China}
 \affiliation{Origin Quantum Computing, Hefei, Anhui 230026, P. R. China}


\begin{abstract}
  While quantum computing provides an exponential advantage in solving system of linear equations, there is little work to solve system of nonlinear equations with quantum computing. 
  We propose quantum Newton’s method (QNM) for solving $N$-dimensional system of nonlinear equations based on Newton's method.
  In QNM, we solve the system of linear equations in each iteration of Newton's method with quantum linear system solver. We use a specific quantum data structure and $l_{\infty}$ tomography with sample error $\epsilon_s$ to 
  implement the classical-quantum data conversion process between the two iterations of QNM,
  thereby constructing the whole process of QNM. 
  The complexity of QNM in each iteration is $O(\log^4N/\epsilon_s^2)$. Through numerical simulation, we find that when $\epsilon_s>>1/\sqrt{N}$, QNM is still effective, so the complexity of QNM is sublinear with $N$, which provides quantum advantage compared with the optimal classical algorithm.

\end{abstract}


\maketitle

\section{Introduction}
The system of nonlinear equations appears in many fields, such as nonlinear finite element analysis\cite{reddy2004nonlinear}, nonlinear dynamics\cite{thompson2002nonlinear}, nonlinear programming\cite{mangasarian1994nonlinear}, etc. In general, the analytical solutions of system of nonlinear equations cannot be obtained effectively. Numerical methods are often used to solve system of nonlinear equations.
Newton's method is a basic method for solving system of nonlinear equations\cite{rheinboldt1974methods,burden_numerical_2016}. 
However, the complexity of solving a system of nonlinear equations with Newton's method grows polynomially with the dimension of the equations\cite{burden_numerical_2016}, using Newton's method to solve high-dimensional nonlinear equations is intractable. It is important to develop more efficient algorithms for solving system of nonlinear equations.

Quantum computing provides a promising way to speed up the solution of system of nonlinear equations. In recent years there have been many quantum algorithms developed to solve various equations, such as system of linear equations\cite{harrow2009quantum,childs2017quantum,Suba2019quantum}, system of linear differential equations\cite{Clader_2013,berry2014highorder,Montanaro_2016,berry2017quantum,xin2020quantum,cao2013quantum,costa2019quantum,fillion2017quantum,engel2019quantum,arrazola2019quantum,linden2020quantum,childs2020quantumspectral,childs2020highprecision} and nonlinear differential equations\cite{leyton2008quantum,lubasch2020variational,liu2020efficient,lloyd2020quantum}. 

However, there is little work for the solution of system of nonlinear equations. Qian $et\ al$.\cite{qian2019quantum} proposed a quantum algorithm for solving system of nonlinear equations, their work is based on Grover algorithm and only has square acceleration. Rebentrost $et\ al$.\cite{rebentrost2019quantum} proposed a quantum Newton's method for constrained polynomial optimization, the complexity of their method grows exponentially as iteration times increases.

In our work, we propose a quantum Newton’s method (QNM) for the solution of system of nonlinear equations. QNM is an iteration method, in each iteration, we solve a system of linear equations with quantum linear system solver (QLSS). Between two iterations, there are some classical-quantum data conversions. We realize the conversions by using a quantum data structure and $l_{\infty}$ tomography\cite{kerenidis_quantum_2019} with sample error $\epsilon_s$.

The complexity of QNM in each iteration is $O(\log^4N/\epsilon_s^2)$. Through numerical simulation, we found that when $\epsilon_s>>1/\sqrt{N}$, QNM is still effective, so the complexity of QNM is sublinear with $N$. Compared with the best classical algorithm, QNM has a significant acceleration. The complexity of QNM grows linearly with iteration times, which is better than the work proposed in \cite{rebentrost2019quantum}.

This paper is organized as follows. We introduce Newton's method in Sec.~\ref{newton}. Then the details of QNM are discussed in Sec.~\ref{Section2} and we analyze the complexity of QNM in Sec.~\ref{complexity}. Sec.~\ref{appl} gives some applications of QNM and some numerical simulation to verify the effectiveness of QNM. Finally, we summarize our work and propose some future research directions in Sec.~\ref{conclusion}.

\section{Newton's Method}\label{newton}

Newton's method is an iterative method for solving system of nonlinear equations. It obtains solution of the system of nonlinear equations by iteratively solving system of linear equations.

The system of nonlinear equations is defined as:
\begin{equation}\label{nonlinear_eq}
  \bm F(\bm x)=\bm 0,
\end{equation}
where
\begin{equation}
  \bm F(\bm x)=
  \begin{bmatrix}
      f_1(\bm x) \\
      \vdots  \\
      f_N(\bm x)
  \end{bmatrix},
  \bm x=
  \begin{bmatrix}
      x_1 \\
      \vdots \\
      x_N
  \end{bmatrix},
  \bm 0=
  \begin{bmatrix}
      0 \\
      \vdots \\
      0
  \end{bmatrix},
\end{equation}
each $f_i(\bm x)$ is a real-valued nonlinear function that maps $\bm x\in\mathbb{R}^N$ to $\mathbb{R}$. We use $\bm x^*$ to represent the solution of Eq.(\ref{nonlinear_eq}). The Jacobi matrix of $\bm F(\bm x)$ is defined as:
\begin{equation}\label{jacobi_matrix}
  \bm F'(\bm x)=(\partial_jf_i(\bm x))_{N\times N}=(a_{ij})_{N\times N},
\end{equation}
The process of Newton's method is shown in Algorithm $\bm{1}$.

\begin{algorithm}[H]
  \label{alg:NewtonIteration}
  \caption{ Newton's method} 
  \begin{flushleft}
    $\ ${\bf Input:} $\bm F(\bm x)$, initial vector $\bm x^0$ and accuracy error $\epsilon$.
  \end{flushleft}

  \begin{flushleft}
    $\ ${\bf Output:} Solution $\bm x^*$.
  \end{flushleft}
  \begin{algorithmic}[1] 
    \STATE Assuming $k$ iterations have been performed, $\bm x^k$ and $F(\bm x^k)$ have been found. Compute Jacobi matrix $\bm F'(\bm x^k)=A_k$, and define $\bm b_k=-\bm F(\bm x^k)$;
    \label{newton:2}
    \STATE Solve the system linear of equations:
    \begin{equation}\label{eq_newton_linear}
    A_k\Delta\bm x^k=\bm b^k.
    \end{equation}
    \STATE Compute $\bm x^{k+1}=\bm x^k+\Delta\bm x^k$ and $\bm F(\bm x^{k+1})$.
    \STATE If $||\bm F(\bm x^{k+1})||\leq \epsilon$, we have $\bm x^*=\bm x^{k+1}$
    and go to step \ref{newton:end}.
    Otherwise, $k+1\to k, \bm x^{k+1}\to \bm x^k, \bm F(\bm x^{k+1})\to \bm F(\bm x^k)$, go to step \ref{newton:2}.
    \STATE Return $\bm x^*$.
    \label{newton:end}
    \STATE End.
  \end{algorithmic}
\end{algorithm}

\section{QNM} \label{Section2}

In our work, we only consider sparse system of nonlinear equations, the Jacobi matrix $\bm F'(\bm x)$ is a $d$-sparse matrix, which means $\bm F'(\bm x)$ has at most $d$ non-zero elements in any row or column. Throughout this paper, $d$ represents the sparsity of $\bm F'(\bm x)$. In QNM, we need some oracles of Eq.(\ref{nonlinear_eq}).
Suppose we are given oracles $O_{f1}$ and $O_{f2}$ that provide the information of Eq.(\ref{nonlinear_eq}), $O_{f1}$ and $O_{f2}$ are defined as:
\begin{equation}
  O_{f1}|i\rangle|j\rangle=|i,f(i,j)\rangle, i=1,2,...N,j=1,2,...d
\end{equation}
\begin{equation}
  O_{f2}|i\rangle|\bm x^{(i)}\rangle|0\rangle=|i\rangle|\bm x^{(i)}\rangle|f_i(\bm x^{(i)})\rangle, i=1,2,...N.
\end{equation} 
where $f(i,j)$ is the subscript of the $j$-th variable in $f_i(\bm x)$, and $\bm x^{(i)}$ represents the related $x_j$ in $f_i(\bm x)$. The construction of $O_{f1}$ and $O_{f2}$ are related to specific problems. In many specific problems, $O_{f1}$ and $O_{f2}$ can be implemented with quantum arithmetic. In Sec.\ref{test_problem}, we give some specific examples.

\begin{figure}
  \centerline{
  \includegraphics[width=0.40\textwidth]{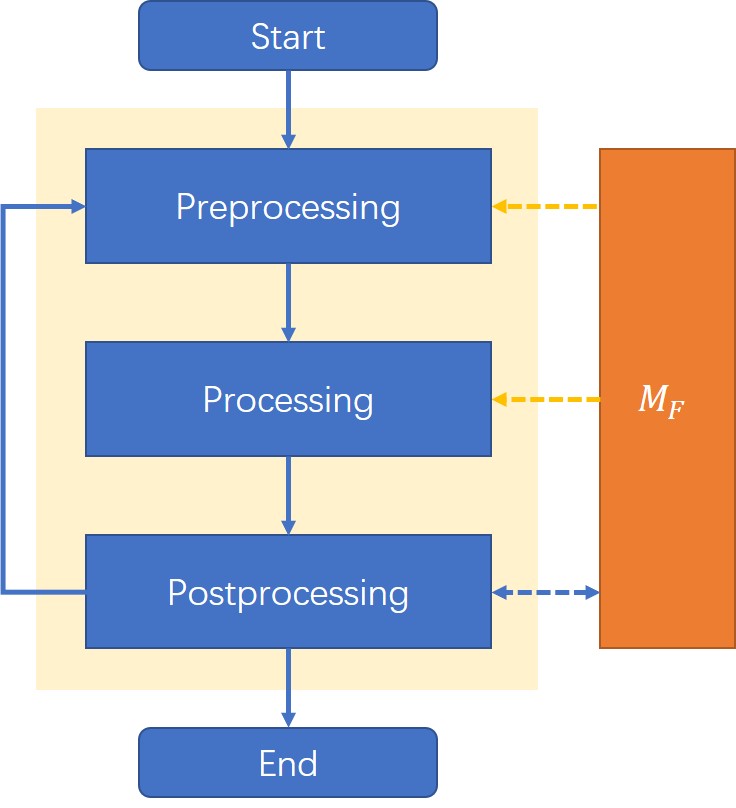}}
  \caption{Overall block of quantum Newton's method.} \label{fig-1a}
\end{figure}

QNM is a quantum version of Newton's method. In each iteration of Newton’s method, we solve a system of linear equations represented by Eq.(\ref{eq_newton_linear}). In QNM, we implement the process of solving the system of linear equations with QLSS, and give the conversion process of classical-quantum data during each iteration. 
The overall block diagram of QNM is shown in Fig.\ref{fig-1a}. Each iteration of QNM is divided into three steps: preprocessing, processing and postprocessing. The details of these three steps are introduced in Sec.\ref{preprocess}, Sec.\ref{qlsa} and Sec.\ref{post_processing}.
During the execution of these 3 steps, we use a quantum data structure $M_F$. Let us introduce $M_F$ first.

\begin{figure}
  \centerline{
  \includegraphics[width=0.40\textwidth]{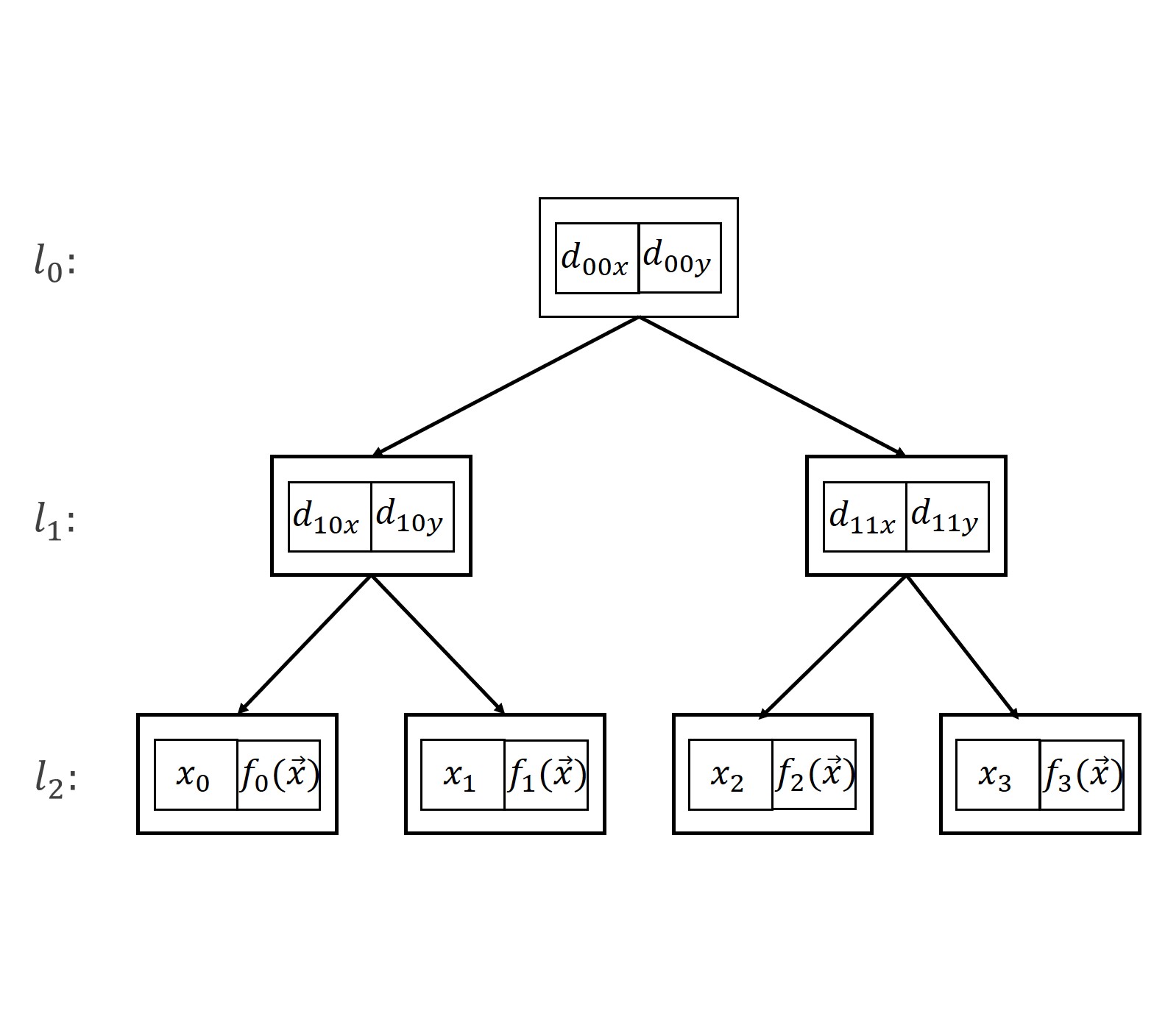}}
  \caption{Structure of a 4-dimensional $M_F$.} \label{fig-1b}
\end{figure}

\subsection{Quantum data structure}\label{oracle}

In QNM, we use $M_F$ to implement some sub-processes. The structure of $M_F$ is inspired by previous work\cite{kerenidis_quantum_2016,giovannetti2008quantum,giovannetti2008architectures}. $M_F$ is a binary tree structure. In each leaf node, we store $x_i$ and $f_i(\bm x)$, In other nodes, we store $d_{ijx}$ and $d_{ijy}$, which are defined as:
\begin{align}
  &d_{ijx}=\sqrt{\sum_{k=j\times 2^{n-i}}^{(j+1)\times 2^{n-i}-1}{x_k^2}},d_{ijy}=\sqrt{\sum_{k=j\times 2^{n-i}}^{(j+1)\times 2^{n-i}-1}{f_k^2(\bm x)}}\notag\\
  &i=0,...,n-1,\ j=0,...,2^{n-i}-1,
\end{align}
where $i,j$ represent the $j$-th node of the $i$-th layer, $n=\lceil\log N\rceil$. In the root node, we store $d_{00x}=||\bm x||$ and $d_{00y}=||\bm F(\bm x)||$.
Fig.~\ref{fig-1b} shows the structure of $M_F$ with dimension $N=4$. Given an $N$-dimensional $M_F$, it has such properties\cite{kerenidis_quantum_2016,giovannetti2008quantum,giovannetti2008architectures}:
\begin{itemize}
  \item [(1)] Extract $x_i$ or $f_i(\bm x)$ in $O(\log^2N)$ time complexity, which are represented with $O_{M1}$ and $O_{M2}$:
  \begin{align}\label{om12}
    &O_{M1}|i\rangle|0\rangle=|i\rangle| x_i\rangle,\notag\\ 
    &O_{M2}|i\rangle|0\rangle=|i\rangle|f_i(\bm x)\rangle,\notag\\ 
    &i=0,1,...,N-1.
  \end{align}
  \item [(2)] Extract the norm of part of $\bm x$ or $\bm F(\bm x)$ in $O(\log^2N)$ time complexity, which are represented with $O_{M3}$ and $O_{M4}$:
  \begin{align}\label{om34}
    &O_{M3}|i\rangle|j\rangle|0\rangle=|i\rangle|j\rangle|d_{ijx}\rangle,\notag\\
    &O_{M4}|i\rangle|j\rangle|0\rangle=|i\rangle|j\rangle|d_{ijy}\rangle,\notag\\
    &i=0,1,...,n,\ j=0,1,...,2^i-1.
  \end{align}
  \item [(3)] The time complexity of Updating the whole $M_F$ is $O(M\log^2 N)$, where $M$ is the sum of the number of $x_i$ and $f_j(\bm x)$ to be updated.
\end{itemize}

\subsection{Preprocessing}\label{preprocess}

Now we introduce three steps of each iteration in QNM. In preprocessing step, we build operations to extract the information of the system of linear equations in this iteration, which is defined in Eq.(\ref{linear_eq}):
\begin{equation}\label{linear_eq}
  A\Delta \bm x=-|\bm b\rangle,
\end{equation}
$A$ and $|\bm b\rangle$ are defined as:
\begin{equation}\label{matrixA}
  A=\frac{\bm F'(\bm x)}{||\bm F'(\bm x)||_{max}},
\end{equation}
\begin{equation}\label{norm_b}
  |\bm b\rangle=\frac{1}{C_b}\sum_i{f_i(\bm x)|i\rangle},
\end{equation}
where $C_b$ is the normalization constant of $\bm F(\bm x)$, $||\bm F'(\bm x)||_{max}:=\max_{j,k}|F'(\bm x)_{jk}|$. We assume that an upper bound of $||\bm F'(\bm x)||_{max}$ can be estimated and we regard the upper bound as $||\bm F'(\bm x)||_{max}$. 

We build $O_b$ to prepare $|\bm b\rangle$ and $O_{A1}$, $O_{A2}$ to extract the information of matrix $A$. 
$O_b$ is defined as:
\begin{equation}
  O_b|\bm 0\rangle=|\bm b\rangle.
\end{equation}
We use the method proposed in \cite{grover2002creating} to build $O_b$. In $O_b$'s construction process, we query $O_{M4}$ $O(\log N)$ times.
$O_{A1}$ and $O_{A2}$ are defined as:
\begin{equation}\label{oa1}
  O_{A1}|j,l\rangle=|j,h(j,l)\rangle,
\end{equation}
\begin{equation}\label{oa2}
  O_{A2}|j,k,z\rangle=|j,k,z\oplus A_{jk}\rangle,
\end{equation}
where $h(j,l)$ is the column number of A's $l$-th non-zero element. It is easy to see that $O_{A1}$ and $O_{f1}$ are the same. We only need to build $O_{A2}$. For simplicity, we set $|z\rangle=|0\rangle$, $O_{A2}$ is rewritten as:
\begin{equation}\label{rOa2}
  O_{A2}|j,k,0\rangle=|j,k,||\bm F'(\bm x)||_{max}^{-1}\frac{\partial{f_j(\bm x)}}{\partial{x_k}}\rangle
\end{equation}
To build $O_{A2}$, we need a quantum circuit to compute $\frac{\partial{f_j(\bm x)}}{\partial{x_k}}$. We use finite difference method (FDM) to approximate $\frac{\partial{f_j(\bm x)}}{\partial{x_k}}$. For example, Eq.(\ref{dif_format}) gives an approximation of $\frac{\partial{f_j(\bm x)}}{\partial{x_k}}$.
\begin{equation}\label{dif_format}
  \frac{\partial{f_j(\bm x)}}{\partial{x_k}}\approx\frac{f_j(\bm x+\Delta x_k)-f_j(\bm x)}{\Delta x_k}
\end{equation}
For each $f_j(\bm x)$, we extract $\bm x^{(j)}$ by querying $O_{f1}$ and $O_{M1}$ $d$ times. Next we use $O_{f2}$ to compute $|f_j(\bm x)\rangle$. We can also change $|\bm x^{(j)}\rangle$ to $|\bm x^{(j)}+\Delta x_k\rangle$ and compute $|f_j(\bm x^{(j)}+\Delta x_k)\rangle$. Finally we compute $A_{jk}$ and uncompute the ancilla qubits we used. The whole process is shown as follows:
\begin{align*}
  &|j,k\rangle|0\rangle|0\rangle\\
  \xrightarrow{O_{f1},\ O_{M1}} & |j,k\rangle|\bm x^{(j)}\rangle|0\rangle \\
  \xrightarrow{O_{f2},\ FDM} & |j,k\rangle|\bm x^{(j)}\rangle|\frac{\partial{f_j(\bm x)}}{\partial{x_k}}\rangle \\
  \xrightarrow{uncompute\quad \bm x^{(j)}} & |j,k\rangle|0\rangle|\frac{\partial{f_j(\bm x)}}{\partial{x_k}}\rangle \\
  \to & |j,k\rangle|0\rangle|A_{jk}\rangle
\end{align*}

\subsection{Processing}\label{qlsa}

In processing step, we use QLSS proposed in \cite{childs2017quantum} to solve Eq.(\ref{linear_eq}). We input $O_b$, $O_{A1}$ and $O_{A2}$ and get $|\Delta \bm x\rangle$, the normalized solution of Eq.(\ref{linear_eq}). The success rate of the QLSS is
\begin{equation}\label{p_value}
  p=\frac{||A^{-1}|b\rangle||_2^2}{\alpha^2},
\end{equation}
$\alpha$ is defined as:
\begin{equation}\label{alpha}
  \alpha:=\frac{4}{d}\sum_{j=0}^{j_0}
  \frac{\sum_{i=j+1}^{c}\begin{pmatrix}
  2c\\c+i
  \end{pmatrix}}{2^{2c}},
\end{equation} 
where $j_0=\sqrt{c \log(4c/\epsilon)}$ and $c=\kappa^2\log(\kappa/\epsilon)$\cite{childs2017quantum}. $p$ can be estimated with quantum amplitude estimation algorithm\cite{brassard2000quantum}.

We execute QLSS multiple times and use $l_\infty$ tomography\cite{kerenidis_quantum_2019} to sample output state $|\Delta \bm x\rangle$ and get the sampled state $|\widetilde{\Delta \bm x\rangle}$. The $l_\infty$ tomography is shown in theorem \ref{sample_theorem}, the sample times is $O(\frac{\log N}{\epsilon_s^2})$, where $\epsilon_s$ represents the sample error. We also compute the normalization constant of $\Delta \bm x$: $C_{\Delta \bm x}=||\Delta \bm x||_2$.
Combine Eq.(\ref{linear_eq}), Eq.(\ref{matrixA}), Eq.(\ref{norm_b}) and Eq.(\ref{p_value}), we have
\begin{equation}\label{c_delta_x}
  C_{\Delta \bm x}=\frac{\alpha C_b\sqrt{p}}{||\bm F'(\bm x)||_{max}}.
\end{equation}
\begin{Theo}
  \label{sample_theorem}
  ($l_\infty$ tomography\cite{kerenidis_quantum_2019}). 
  Given access to unitary $U$ such that $U|0\rangle=|x\rangle$ and its controlled version in time $T(U)$, there is a tomography algorithm with time complexity
  $O(T(U)\frac{\log N}{\epsilon_s^2})$ that produces unit vector $\widetilde{X}\in R^N$ such that $||\widetilde{X}-x||_{\infty}\leq \epsilon_s$ with probability at least $(1-1/poly(N))$.
\end{Theo}

\subsection{Postprocessing}\label{post_processing}

In postprocessing step, we use the sampled state $|\widetilde{\Delta \bm x\rangle}$ to update $M_F$ and determine whether to stop the iteration. 

We first update the root node of $M_F$: $x_i$ and $f_i(\bm x)$. 
The update of $x_i$ is:
\begin{equation}\label{mk_update}
  |\bm x\rangle \to |\bm x + C_{\Delta \bm x}\widetilde{\Delta \bm x}\rangle\approx|\bm x-\bm F'^{-1}(\bm x)\bm F(\bm x)\rangle
\end{equation}
From $|\widetilde{\Delta \bm x\rangle}$ we compute the changed $f_i(\bm x)$ and update the changed $f_i(\bm x)$. we use $N_x$, $N_f$ to represent the number of changed  $x_i$ and $f_i(\bm x)$ respectively. $N_x$, $N_f$ satisfy:
\begin{equation}
  N_x<\frac{\log N}{\epsilon_s^2},\ N_f<d\frac{\log N}{\epsilon_s^2}.
\end{equation}
Then the update complexity of the whole $M_F$ is $O(d\frac{\log^3 N}{\epsilon_s^2})$.

The iteration cutoff condition is $||\bm F(\bm x)||<\epsilon$. At the end of each iteration, we use $O_{M4}$ introduced in Eq.(\ref{om34}) to compute $||\bm F(\bm x)||$, we set $i=0,\ j=0$ and have
\begin{equation}
  O_{M4}|0\rangle|0\rangle|0\rangle=|0\rangle|0\rangle|||\bm F(\bm x)||\rangle,
\end{equation}
If $||\bm F(\bm x)||<\epsilon$, we stop iteration, otherwise, we execute the next iteration until the iteration stops.

Up to now, we have introduced the whole process of QNM. The process is also shown in Algorithm $\bm 2$. The output of QNM is the updated $M_F$, it saves the solution of the system of nonlinear equations.

\begin{algorithm}[H]
  \caption{ Quantum Newton's method}
  \begin{flushleft}
    $\ ${\bf Input:}
    initial $M_F$,$O_{f1}$, $O_{f2}$, sample error $\epsilon_s$ and  accuracy error $\epsilon$.
  \end{flushleft}

  \begin{flushleft}
    $\ ${\bf Output:}
    $M_F$.
  \end{flushleft} 
  \begin{algorithmic}[1] 
  \STATE Construct $O_b$, $O_{A1}$ and $O_{A2}$.
  \label{qnewton:2}
  \STATE Input $O_b$, $O_{A1}$ and $O_{A2}$ into QLSS introduced in Sec.\ref{qlsa}, execute QLSS $O(\frac{\log N}{\epsilon_s^2})$ times and get $|\widetilde{\Delta \bm x}\rangle$ with $l_\infty$ tomography.
  \STATE Compute $C_{\Delta \bm x}$ with Eq.(\ref{c_delta_x}). 
  \STATE Update $M_F$ with the process decribed in Sec.\ref{post_processing}.
  \STATE Compute $||\bm F(\bm x)||$, if $||\bm F(\bm x)||\leq \epsilon$, turn to step \ref{qnewton:end}.
  Otherwise, turn to step \ref{qnewton:2}.
  \STATE Return $M_F$.
  \label{qnewton:end}
  \STATE End.
  \end{algorithmic}
  \label{alg:QuantumNewtonIteration}
\end{algorithm}

\section{Complexity}\label{complexity}

In this section, we analyze the query complexity and time complexity of QNM. First, we analyze the query complexity. 

\subsection{Query complexity}

In QNM, we use $O_{f1}$, $O_{f2}$ and $M_F$ to build other operations. Now we analyze the query complexity of $O_{f1}$, $O_{f2}$ and $M_F$.

In $O_b$'s construction process, we query $M_F$ $O(\log N)$ times\cite{grover2002creating}. In $O_{A2}$'s construction process introduced in Sec.\ref{preprocess}, we query $O_{f1}$, $M_F$ $O(d)$ times and query $O_{f2}$ $O(1)$ times. 

In QLSS, improved by variable-time amplitude amplification\cite{ambainis2010variable}, the query complexity of $O_{A1}$, $O_{A2}$ and $O_b$ is $O(d\kappa \text{poly}(\log(d\kappa/\epsilon)))$\cite{childs2017quantum}, $\kappa$ represents the condition number of matrix $A$. As shown in Theorem.\ref{sample_theorem}, the sampling complexity is $O(\frac{\log N}{\epsilon_s^2})$. In postprocessing step, we query $M_F$ $O(1)$ times and we can omit it.
Therefore in each iteration, the query complexity of $M_F$ is
\begin{equation}
O((d+\log N)\frac{d\kappa\log N}{\epsilon_s^2}\text{poly}(\log(d\kappa/\epsilon))),
\end{equation}
the query complexity of $O_{f1}$ is
\begin{equation}
  O(\frac{d^2\kappa\log N}{\epsilon_s^2}\text{poly}(\log(d\kappa/\epsilon))),
\end{equation}
and the query complexity of $O_{f2}$ is
\begin{equation}
  O(\frac{d\kappa\log N}{\epsilon_s^2}\text{poly}(\log(d\kappa/\epsilon))).
\end{equation}

\subsection{Time Complexity}

Next, we analyze the time complexity of QNM. Here we regard the time complexity of $O_{f1}$ and $O_{f2}$ as $O(1)$. 
In construction process of $O_b$, we query $O_{M4}$ $\log N$ times and perform $\log N$ quantum arithmetic operations, the time complexity of quantum arithmetic is $O(poly(\log(1/\epsilon)))$\cite{mitarai2019quantum}, therefore the time complexity of $O_b$ is:
\begin{align}
  T(O_b)&=O(\log N\times (poly(\log(1/\epsilon))+ \log^2N))
\end{align}
$O_{A2}$ is built with $O_{f1}$, $O_{f2}$, $O_{M1}$ and quantum arithmetic, the time complexity of $O_{A2}$ is 
\begin{equation}
  T(O_{A2})=O(d\log^2N+poly(\log(1/\epsilon)))
\end{equation}
Then the time complexity of QLSS is\cite{childs2017quantum}
\begin{align}
  &T(QLSS)\notag\\
  =&O(d\kappa^2 \text{poly}(\log(d\kappa/\epsilon))(T(O_b)+T(O_{A1})+T(O_{A2})))\notag\\
  =&O(d\kappa^2 \text{poly}(\log(d\kappa/\epsilon)(\log^3N+d\log^2N))
\end{align}
The $\kappa$-dependence of $T(QLSS)$ can be improved from quadratic to nearly linear with variable-time amplitude amplification\cite{ambainis2010variable,childs2017quantum}, then
\begin{equation}
  T(QLSS)=O(d\kappa\text{poly}(\log(d\kappa/\epsilon)(\log^3N+d\log^2N))
\end{equation}
In $l_{\infty}$ tomography, we execute QLSS $O(\frac{\log N}{\epsilon_s^2})$ times. Finally, the time complexity of updating $M_F$ is $O(\frac{d\log^3 N}{\epsilon_s^2})$.
In summary, the time complexity of one iteration is 
\begin{equation}\label{com_one}
  T_q=O\left(\frac{\log^3 N}{\epsilon_s^2}d\kappa \text{poly}(\log(d\kappa/\epsilon)(\log N+d)\right)
\end{equation}

In each iteration of classical Newton's method, we need to solve a system of linear equations, and one of the best algorithms to solve the system of linear equations is conjugate gradient algorithm\cite{shewchuk1994conjugate}, the complexity of conjugate gradient algorithm is $O(Nd\kappa\log(1/\epsilon))$. the complexity of other processes in classical Newton's method can be ignored. Therefore the time complexity of one iteration of classical Newton's method is 
\begin{equation}\label{com_class}
  T_c=O(Nd\kappa\log(1/\epsilon))
\end{equation}

Compare Eq.(\ref{com_one}) and Eq.(\ref{com_class}), we find that the $\kappa$-dependence and $\epsilon$-dependence are nearly the same. The difference is mainly in $N$, $d$ and $\epsilon_s$. 
In general, most of the real systems are sparse, $d$ has little effect on complexity, we can ignore $d$. The complexity of QNM contains $\epsilon_s$. We do some numerical simulations in Sec.\ref{appl} and find when $\epsilon_s>>1/\sqrt{N}$, QNM is still effective, then the time complexity of QNM is sublinear with $N$. Therefore, compared with classical Newton's method, QNM provides a quantum advantage.

\section{Application and Numerical Simulation}\label{appl}

QNM can be applied to many fields, including computational fluid dynamics, nonlinear dynamics, economics, etc., to solve nonlinear problems in these fields. A general solution process is divided into the following steps: (1) Given a specific nonlinear problem, such as a nonlinear differential equation; (2) Discretize the problem by numerical methods, such as FDM, FEM, etc., to obtain the system of nonlinear equations to be solved. (3) Use QNM to solve the system of nonlinear equations and obtain the solution of the problem.

Compared with classical Newton's method, we have an $l_{\infty}$ tomography in each iteration and update $\bm x$ with sampled state, it will affect iteration speed or precision of QNM. To prove the effectiveness of QNM, we do some numerical simulations to test the influence of sampling error $\epsilon_s$ on QNM.

We choose two problems: (1) Nonlinear first-order diffusion problem; (2) Beam lateral vibration problem. We use FDM to discretize the two problems and use QNM to solve the systems of nonlinear equations obtained by FDM.
We first introduce the details of these two problems.

\subsection{Problems and difference format}\label{test_problem}

\subsubsection{Nonlinear first-order diffusion problem}\label{app11}
The first problem is solving a nonlinear first-order diffusion problem with our method, the problem is defined as: 
\begin{equation}\label{app1}
  \begin{cases}
  \frac{\partial u(x, t)}{\partial t}=g(u,x,t)\frac{\partial u(x, t)}{\partial x}+f(x, t),-1<x<1,t>1,\\
  g(u,x,t)=-1+x \sin (u(x, t)),\\
  f(x,t)=2e^{2t}[x^2-1+x-x^2\sin (e^{2t}(e^2-1))]
  \end{cases}
\end{equation}
The initial condition and boundary condition are:
\begin{equation}
  \begin{cases}
  u(x,0)=x^2-1,-1<x<1;\\
  u(-1,t)=0,t>0.
  \end{cases}
\end{equation}
We discrete the equation in a such way:
\begin{equation}
  \begin{cases}
    x_i=-1+(i+1)h,i=0,1,...,N_1-1,h=\frac{2}{N_1},\\
    t_j=(j+1)s,j=0,1,...,N_2-1,s=\frac{1}{N_2}.
  \end{cases}
\end{equation}
We define $u_{i,j}=u(x_i,t_j)$, $u_{i,j,x}$ and $u_{i,j,t}$ are written as:
\begin{equation}
  \begin{cases}
    u_{i,j,x}=\frac{u_{i,j}-u_{i-1,j}}{h},\\
    u_{i,j,t}=\frac{u_{i,j}-u_{i,j-1}}{s}.
  \end{cases}
\end{equation}
Then we have the following system of nonlinear equations:
\begin{equation}\label{app1_non}
  \begin{cases}
    u_{i,j,t}+(1-x_i\sin (u_{i,j}))u_{i,j,x}-f(x_i,t_j)=0,\\
    i=0,1,...,N_1-1;\ j=0,1,...,N_2-1.\\
    u_{-1,j}=0,u_{i,-1}=x_i^2-1.
  \end{cases}
\end{equation}
The Jacobi Matrix of Eq.(\ref{app1_non}) is an $N_1\times N_2$-dimensional sparse square matrix and the position of non-zero element is regular, the position of non-zero element of $m=N_1\times j+i$ row is represented as:
\begin{equation}\label{appl1_oa1}
  S(m)=
  \begin{cases}
    \{m\},\ i=0,j=0, \\
    \{m-1,m\},\ i\neq0,j=0, \\
    \{m-N_1,m\},\ i=0, j\neq0,\\
    \{m-N_1,m-1,m\},\ i\neq0, j\neq0.
  \end{cases}
\end{equation}
The oracles $O_{f1}$ and $O_{f2}$ of Eq.(\ref{app1_non}) can be constructed by realizing Eq.(\ref{appl1_oa1}) and Eq.(\ref{app1_non}) with some quantum arithmetic circuits. The gate complexity is $O(\log N+\text{poly}(\log (1/\epsilon)))$\cite{nielsen_quantum_2002}.
When considering the complexity of $O_{f1}$ and $O_{f2}$ in the overall time complexity of our method, we can find that the complexity of $O_{f1}$ and $O_{f2}$ can be ignored, and the overall time complexity does not change.

\subsubsection{Beam lateral vibration problem}

The second problem is a beam lateral vibration problem, the problem is defined as:

\begin{equation}\label{app3}
  \begin{cases}
  g(x)\frac{\partial^4\omega}{\partial x^4}+2\frac{\partial g}{\partial x}\frac{\partial^3\omega}{\partial x^3}+\frac{\partial^2 g}{\partial x^2}\frac{\partial^2\omega}{\partial x^2}
  +\mu(x)\frac{\partial^2\omega}{\partial t^2}\\
  +G(\omega)=f(x,t),\ -4\leq x\leq4,0\leq t\leq 2,\\
  \mu(x)=e^{-2x^2}+1,g(x)=2e^{-2x^2}+1,\\
  G(\omega)=(1+2e^{-2x^2})\omega+(5+e^{-3x^2})\omega^3,
  \end{cases}
\end{equation}
where $f(x,t)$ is determined by analytical solution $\omega(x,t)=e^{-x^2-t^2}$. The initial condition and the boundary condition are:
\begin{equation}
  \begin{cases}
    \omega(x,0)=e^{-x^2},\omega_t(x,0)=0,\\
    \omega(\pm 4,t)=0,\omega_x(\pm4,t)=0.
  \end{cases}
\end{equation}
We discrete the equation in a such way:
\begin{equation}
  \begin{cases}
    x_i=-4+(i+1)h,i=0,1,...,N_1-1,h=\frac{8}{N_1},\\
    t_j=(j+1)s,j=0,1,...,N_2-1,s=\frac{2}{N_2}.
  \end{cases}
\end{equation}
$\omega_{i,j}$ represents $\omega(x_i,t_j)$, the difference format is:
\begin{equation}
  \begin{cases}
    \omega_{i,j,xx}=\frac{\omega_{i+1,j}-2\omega_{i,j}+\omega_{i-1,j}}{h^2},\\
    \omega_{i,j,tt}=\frac{\omega_{i,j}-2\psi_{i,j-1}+\omega_{i,j-2}}{s^2},\\
    \omega_{i,j,xxx}=\frac{\omega_{i+2,j}-2\omega_{i+1,j}+2\omega_{i-1,j}-\psi_{i-2,j}}{2h^3},\\
    \omega_{i,j,xxxx}=\frac{\omega_{i+2,j}-4\omega_{i+1,j}+\omega_{i,j}-4\omega_{i-1,j}+\omega_{i-2,j}}{h^4}.
  \end{cases}
\end{equation}
The system of nonlinear equations is represented as:

\begin{equation}\label{app3_non}
  \begin{cases}
    g(x_i)\omega_{i,j,xxxx}+2g_x(x_i)\omega_{i,j,xxx}+g_xx(x_i)\omega_{i,j,xx}\\
    +\mu(x_i)\omega_{i,j,tt}+G(\omega_{i,j})-f(x_i,t_j)=0,\\
    i=0,1,...,N_1-1;\ j=0,1,...,N_2-1.
  \end{cases}
\end{equation}
and based on the boundary condition, we have
\begin{equation}\label{app3_non1}
  \begin{cases}
    \omega_{-1,j}=\omega_{N_1,j}=0,\\
    \omega_{-2,j}=\omega_{0,j},\omega_{N_1+1,j}=\omega_{N_1-1,j},\\
    \omega_{i,-1}=e^{-x_i^2},\omega_{i,-2}=\omega_{i,0}.
  \end{cases}
\end{equation}
Combine Eq.(\ref{app3_non}) and Eq.(\ref{app3_non1}), we have an $N_1\times N_2$-dimensional system of nonlinear equations. Similar to the nonlinear first-order diffusion problem described in Sec.\ref{app11},
the oracles $O_{f1}$ and $O_{f2}$ of this problem can also be constructed by realizing Eq.(\ref{app3_non1}) and Eq.(\ref{app3_non}) with some quantum arithmetic circuits.

\subsection{Numerical simulations}

Now we show the details of our numerical simulations. 
We use C++ to implement our numerical simulations. In our simulation, we use the LU decomposition linear solver\cite{saad2003iterative} in the C++ $Eigen$  template library to solve the linear system in each iteration of QNM and get $\Delta \bm{x}$, then we execute Algorithm $\bm{3}$ in \cite{kerenidis_quantum_2019} to get the sampled $\Delta \bm{x}$ and update $\bm{x}$ with the sampled $\Delta \bm{x}$. 
We set the difference points of the two dimensions as $N_1=200$, $N_2=200$, then we have $N=40000$-dimensional system of nonlinear equations. 

\begin{figure*}[htbp]
  \centering
  \subfigure[\label{fig-3a}]{\includegraphics[width=0.45\textwidth]{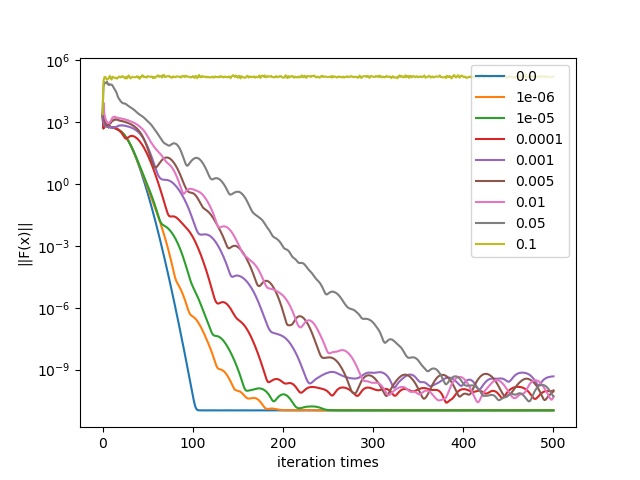}}
  \subfigure[\label{fig-3b}]{\includegraphics[width=0.45\textwidth]{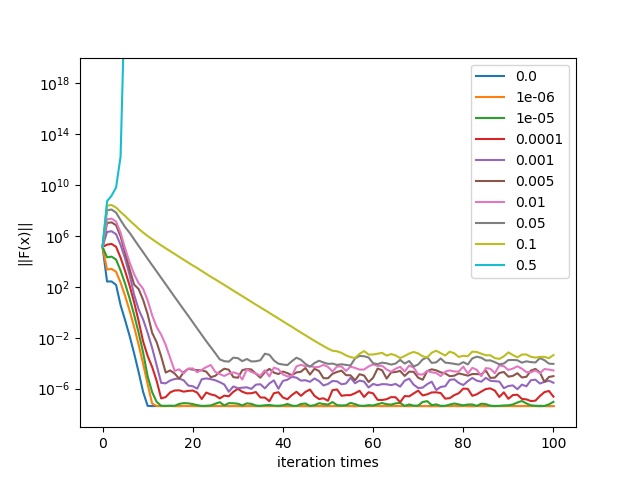}}
  \caption{Numerical Results. Panels (a),(b) represent the influence of sample error $\epsilon_s$ on the convergence process of nonlinear first-order diffusion problem and beam lateral vibration problem respectively.}
  \label{imag_appl1}
\end{figure*}

The numerical simulations of the two problems are shown in Fig.\ref{imag_appl1}.
From the numerical simulations, we get such results: (1) When $\epsilon_s=1/\sqrt{N}=0.005$, QNM converges for the two problems. When $\epsilon_s>1/\sqrt{N}$, QNM also converges, the upper bound of $\epsilon_s$ when QNM converges for the two problems is $0.05$ and $0.1$ respectively.  (2) Sample error $\epsilon_s$ slows down the iteration speed and increases iteration times. Compared with $N$, iteration times has little influence on the complexity of QNM, so we can ignore the influence on iteration times. (3) When $\epsilon_s$ is relatively large, $||\bm F(\bm x)||$ cannot converge to the result that the sampling error is 0, so QNM can only get an approximate solution.

\begin{figure*}[htbp]
  \centering
  \subfigure[\label{fig-4a}]{\includegraphics[width=0.45\textwidth]{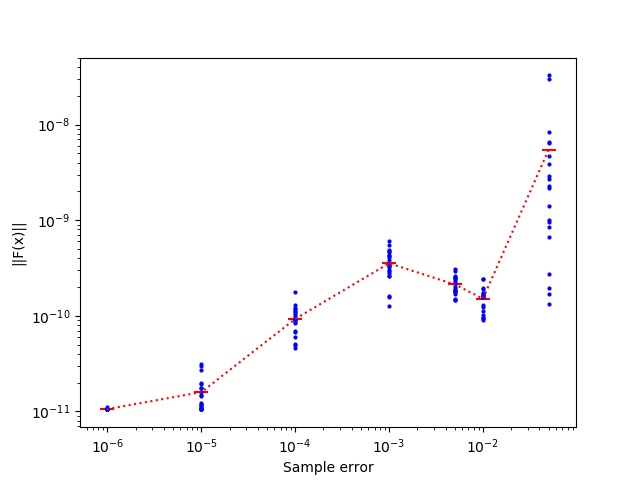}}
  \subfigure[\label{fig-4b}]{\includegraphics[width=0.45\textwidth]{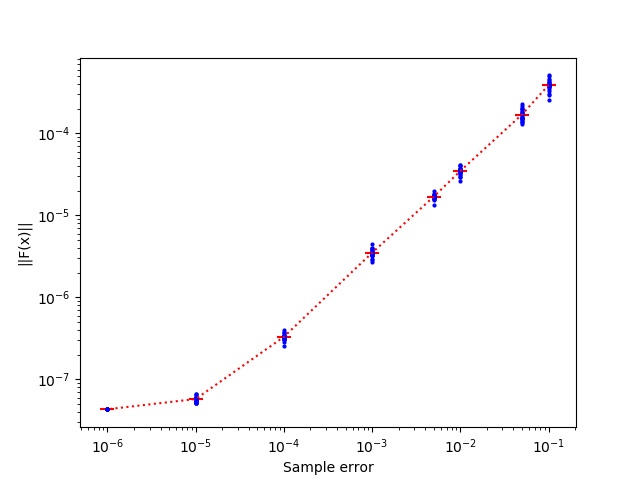}}
  \caption{Relationship between $\epsilon_s$ and convergent $||\bm F(\bm x)||$. Panels (a),(b) represent nonlinear first-order diffusion problem and beam lateral vibration problem respectively.}
  \label{imag_appl2}
\end{figure*}

We also test the effects of $\epsilon_s$ on convergent $||\bm F(\bm x)||$. In Fig.\ref{imag_appl2}, we repeat QNM 20 times to get the statistical results of the effects of $\epsilon_s$ on convergent $||\bm F(\bm x)||$. 
We get the following results: (1) As $\epsilon_s$ increases, the convergent $||\bm F(\bm x)||$ also increases. In Fig.\ref{fig-4a}, there is an interval in which $\epsilon_s$ increases and convergent $||\bm F(\bm x)||$ decreases, but the overall trend that convergent $||\bm F(\bm x)||$ increases with the increase of $\epsilon_s$ remains unchanged. (2) In both problems, convergent $||\bm F(\bm x)||<<\epsilon_s$, so even if we take a relatively large $\epsilon_s$, we can still get a result error much lower than $\epsilon_s$.

From these numerical simulations, we can see that when $\epsilon_s>>1/\sqrt{N}$, QNM is still effective, the convergent $||\bm F(\bm x)||$ is much smaller than $\epsilon_s$.

\section{Conclusion}\label{conclusion}

In this paper, we developed QNM for solving system of nonlinear equations. We give the details of the QNM, analyze the query complexity and time complexity of QNM. We also discuss the application of QNM do some numerical simulations to study the influence of $\epsilon_s$ on QNM. We find when $\epsilon_s>>1/\sqrt{N}$, QNM is still effective, so the complexity of QNM is sublinear with $N$. Compared with optimal classical algorithm, QNM provides a quantum advantage.
We also notice that when $\epsilon_s<1/\sqrt{N}$, the complexity of QNM is superlinear with $N$, which is worse than classical Newton's method.

QNM has the potential to provide quantum acceleration in many fields, such as nonlinear finite element analysis\cite{reddy2004nonlinear}, nonlinear dynamics\cite{thompson2002nonlinear}, nonlinear programming\cite{mangasarian1994nonlinear} and so on. In the future, we will develop the application of QNM in various practical problems, test the impact of $\epsilon_s$ on QNM in various specific problems.

\section*{Acknowledgement}

This work was supported by the National Key Research and Development Program of China (Grant No. 2016YFA0301700), the National Natural Science Foundation of China (Grants Nos. 11625419), the Strategic Priority Research Program of the Chinese Academy of Sciences (Grant No. XDB24030600), and the Anhui Initiative in Quantum Information Technologies (Grants No. AHY080000). 

\bibliography{quantumNM}

\end{document}